\documentclass[11pt]{article} 
\usepackage{hyperref}
\usepackage[margin=1in]{geometry}
\usepackage[round]{natbib}
\usepackage{amsbsy}
\usepackage{amsmath}    
\usepackage{amssymb}
\usepackage[figuresright]{rotating}
\usepackage{booktabs}
\newtheorem{theorem}{Theorem}
\usepackage{geometry}
\providecommand{\keywords}[1]{\textbf{\textit{Keywords---}} #1}

\begin{document}

\title{Testing the Mean Matrix in High-Dimensional Transposable Data}
\author{
Anestis Touloumis\\ 
Cancer Research UK Cambridge Institute\\
 University of Cambridge\\
 Cambridge CB2 0RE, U.K.\\
\texttt{Anestis.Touloumis@cruk.cam.ac.uk} \and 
Simon Tavar\'e\\ 
CRUK Cambridge Institute, University of Cambridge\\
 \and John C. Marioni\\
The EMBL-European Bioinformatics Institute
}
\date{}
\maketitle

\begin{abstract}
The structural information in high-dimensional transposable data allows us to write the data recorded for each subject in a matrix such that both the rows and the columns correspond to variables of interest. One important problem is to test the null hypothesis that the mean matrix has a particular structure without ignoring the dependence structure among and/or between the row and column variables. To address this, we develop a generic and computationally inexpensive nonparametric testing procedure to assess the hypothesis that, in each predefined subset of columns (rows), the column (row) mean vector remains constant. In simulation studies, the proposed testing procedure seems to have good performance and, unlike simple practical approaches, it preserves the nominal size and remains powerful even if the row and/or column variables are not independent. Finally, we illustrate the use of the proposed methodology via two empirical examples from gene expression microarrays.
\end{abstract}


\section{Introduction} \label{s:intro}
In some applications, the measurements related to each subject are naturally organized in a matrix, especially when the rows and columns correspond to two different sets of variables and dependencies are expected to occur between and/or among them. \cite{Allen2010} introduced the term `transposable data' to acknowledge the structural information and the presence of two-way dependencies in matrix-valued random variables. Examples of transposable data can be found in spatiotemporal studies \citep{Genton2007,MARDIA1993}, in cross-classified multivariate data \citep{Galecki1994,Naik2001}, in genetics \citep{Allen2010,Allen2012,Efron2009,Teng2009,Yin2012,Ning2011}, in functional MRI \citep{Allen2010}, in time-series \citep{Carvalho2007,Lee2013} and in electroencephalography studies \citep{Zhang1995} among others.

\indent Although our findings can be applied to any of the disciplines mentioned above, our work is primarily motivated by biological studies that use microarrays to study gene expression patterns in multiple tissue samples taken from the same subject \citep{Sottoriva,Zahn2007}. For each subject, the row variables correspond to genes, the column variables to tissue samples and the measurements are mRNA gene expression levels. A complex and high-dimensional dependence structure is expected to occur as neither the genes nor the tissue samples are likely to be independent. In such studies, a natural biological objective is to determine whether given subsets of tissue samples share a common mean vector of gene expression levels. This leads to two important statistical challenges. First, the number of genes will typically exceed the number of subjects and it is a well known fact that classical multivariate tests for testing equality of mean vectors, such as the Hotelling's $T^2$ or Wilk's $\Lambda$, are not applicable in `large $p$, small $N$' settings. Second, the dependence among the tissue samples for each subject might restrict us from utilizing practical approaches that rely on mixing univariate standard testing procedures and multiple testing correction methods. This includes, for example, the approach of testing the significance of each gene across tissue samples based on an analysis of variance (ANOVA) test and adjusting the corresponding $p$-values for multiple testing. This approach requires tissue-wise (column-wise) independence, a rather strong assumption that is unlikely to be met in real datasets.

\indent  To introduce these concepts in mathematical terms, suppose that an experimentalist collects $N$ independent and identically distributed (i.i.d.) transposable $r \times c$ random matrices $\mathbf X_1, \ldots, \mathbf X_N$. For each subject, there are $r$ row variables and $c$ column variables and the high-dimensional setting is indicated by letting the sample size ($N$) be much smaller than the number of observations ($rc$) for a single subject. The goal is to perform hypothesis testing for $\mathbf M=\mathrm{E}[\mathbf X_i]$, the $r \times c$ mean matrix of the transposable data, while accounting for the two-way dependencies.

\indent To illustrate some difficulties of this task, consider the simple hypothesis
\begin{equation}
\mathrm{H}_0: \mathbf M = \boldsymbol \mu \mathbf 1^{T}_{c} \text{ vs. } \mathrm{H}_1: \mathbf M \neq \boldsymbol \mu \mathbf 1^{T}_{c}, 
\label{onegroup}
\end{equation}
where $\boldsymbol \mu$ is an unknown $r$-variate parameter vector and $\mathbf 1_{s}$ denotes an $s$-variate vector of ones. The null hypothesis suggests that the mean relationship between the row and column variables is completely determined by the row variables. In the motivating examples, $\mathrm{H}_0$ in~(\ref{onegroup}) is consistent with no genes showing differential expression across the multiple tissue samples. To the best of our knowledge, no statistical procedure exists to test hypothesis~(\ref{onegroup}) directly in high-dimensional transposable data unless there are only two dependent column variables $(c=2)$. In this case, the test proposed by \cite{Chen2010} for comparing the mean vector of paired high-dimensional random vectors can be used. To accomplish this, one needs to form the vector of the difference of the two columns for each subject and then test the hypothesis of a zero mean vector. Unfortunately, there is no straightforward way to apply or extend this test when $c>2$. In particular, attempts to do this essentially infer rather than test the mean relationship between the row and column variables. For example, suppose that $\mathbf M= [\boldsymbol \mu,-\boldsymbol \mu,\boldsymbol \mu,-\boldsymbol \mu]$ and consider the following naive algorithm to test hypothesis~(\ref{onegroup}). First, create two groups of column variables, one based on the first two columns and the other based on the last two. Second, for each group create $N$ $r$-variate random vectors by averaging the appropriate columns in each matrix, and then for each subject create the $r$-variate vectors of the difference of the two groups. Thirdly, test hypothesis~(\ref{onegroup}) using the test statistic of \cite{Chen2010} as above. It can be shown that this vector-based test statistic will be powerless since the transformed random vectors will indeed have a zero mean vector. 

\indent By contrast, we propose a simple approach to test hypothesis~(\ref{onegroup}) that overcomes these theoretical problems. In this direction, let $\mathbf P_c=\mathbf I_c - \mathbf J_c/c$ where $\mathbf I_s$ is the identity matrix of size $s$ and $\mathbf J_s$ is the $s \times s$ matrix of ones, and let $\mathrm{tr}(\mathbf A)$ denote the trace operator of the matrix $\mathbf A$. Note that $\mathbf P_{c}$ is a symmetric and idempotent ($\mathbf P^2_{c}= \mathbf P_{c}$) matrix such that $\mathrm{tr}(\mathbf M^{T} \mathbf M \mathbf P_{c})=0$ if and only if $\mathrm{H}_0$ in~(\ref{onegroup}) holds. Since the Frobenius norm, $\mathrm{tr}(\mathbf M^{T} \mathbf M \mathbf P_{c})$, measures deviations from $\mathrm{H}_0$ in~(\ref{onegroup}), it seems meaningful to develop a test statistic based on $\sum_{i \neq j} \mathrm{tr}(\mathbf X_i^{T} \mathbf X_j \mathbf P_{c})/[N(N-1)],$ the unbiased estimator of this norm.
Under rather weak conditions about the two-way dependence structure, illustrated in Section~\ref{CovarianceClass}, this estimator asymptotically follows a normal distribution, and hence, the critical region of the test statistic can be defined under $\mathrm{H}_0$. 

\indent The main contribution of this paper is that we allow testing more complicated hypotheses than hypothesis~(\ref{onegroup}) for the mean matrix. In particular, we consider the hypothesis
\begin{equation}
\mathrm{H}_0: \mathbf M = [\boldsymbol \mu_1 \mathbf 1^{T}_{c_{1}},\boldsymbol \mu_2 \mathbf 1^{T}_{c_{2}},\ldots,\boldsymbol \mu_g \mathbf 1^{T}_{c_g}] \text{ vs. } \mathrm{H}_1: \mathbf M \neq [\boldsymbol \mu_1 \mathbf 1^{T}_{c_{1}},\boldsymbol \mu_2 \mathbf 1^{T}_{c_{2}},\ldots,\boldsymbol \mu_g \mathbf 1^{T}_{c_g}] , 
\label{arbitrarygroups}
\end{equation}
where $c_1,\ldots,c_g$ are positive integers such that $\sum_{q=1}^g c_{q}=c$ with at least one $c_q\geq 2$ and $\boldsymbol \mu_1,\ldots,\boldsymbol \mu_g$ are $g$ unknown $r$-variate parameter vectors. $\mathrm{H}_0$ in~(\ref{arbitrarygroups}) states that in each one of the given $g$ column groups there is no column effect upon the mean of the row variables. Since $g$ is known but arbitrary, the proposed testing procedure is not bounded by the number of column groups or the group size under consideration. For example, hypothesis~(\ref{onegroup}) is a special case of hypothesis~(\ref{arbitrarygroups}) with $g=1$ and $c_1=c$ while the hypothesis that two column variables, say the first two, have a common mean vector is obtained by setting $g=c-1$, $c_1=2$ and $c_2,\ldots,c_g=1$. Similarly to testing hypothesis~(\ref{onegroup}), the proposed test statistic will be based on an asymptotic argument via a pivotal quantity that is the unbiased estimator of the distance of the mean matrix from $\mathrm{H}_0$ in~(\ref{arbitrarygroups}). The proposed testing methodology is a global procedure that produces a single $p$-value for testing $\mathrm{H}_0$ in~(\ref{arbitrarygroups}) and it is not seriously restricted by the presence of dependence structures other than the independence. 

\indent The proposed testing procedure can also be employed to determine the mean relationship between row and column variables in many predefined sets of row variables rather than across all row variables. In the motivating examples, the biological interest might lie in finding gene-sets for which the mean vector of expression levels varies across different tissue samples. This could allow better identification of biological processes that are tissue-specific, thus facilitating their exploration in greater detail. In this case, one needs to test hypothesis~(\ref{onegroup}) for each predefined gene-set and then correct the corresponding $p$-values for multiple testing. We illustrate how to perform this type of analysis in Section~\ref{GBexample}.   

\indent The rest of this article is structured as follows. In Section~\ref{TestStatistic}, we introduce the high-dimensional working framework and we construct the test statistic for testing hypothesis~(\ref{arbitrarygroups}). We also discuss the asymptotic power of the proposed test, we argue that the required assumptions that make the high-dimensional setting manageable are weak, we make general comments about practical aspects of the testing procedure and we provide guidelines about how to adjust the proposed methodology to test hypotheses other than hypothesis~(\ref{arbitrarygroups}). In Section~\ref{Simulation}, we examine the performance of the proposed testing methodology in finite samples using simulations. In Section~\ref{Example}, we apply the proposed testing methodology to two microarrays studies where gene expression levels are measured in different tissue samples \citep{Sottoriva,Zahn2007}. In Section~\ref{Discussion}, we summarize the main findings of our research and future research directions.

\section{Test Statistics for the Mean Matrix} \label{TestStatistic}
As the generative process for transposable data, consider a matrix-valued extension of the nonparametric model for vectors considered in \cite{Bai1996} and \cite{Chen2010} 
\begin{equation}
\mathbf X_{i} = \mathbf W_{i} + \mathbf M 
\label{Nonparametricmodel}
\end{equation}
for $i=1,\ldots,N$, where 
\begin{enumerate}
\item $\mathbf M=\mathrm{E}[\mathbf X_i]$ is the $r \times c$ mean matrix,
\item $\mathbf W_i$ is an $r \times c$ matrix of random variables such that $\mathrm{vec}(\mathbf W_i)= \boldsymbol \Sigma^{1/2} \mathrm{vec}(\mathbf Z_i)$, and where $\mathrm{vec}(\mathbf A)$ denotes vectorization of the matrix $\mathbf A$,
\item $\boldsymbol \Sigma=\boldsymbol \Sigma^{1/2} \boldsymbol \Sigma^{1/2}=\mathrm{cov}[\mathrm{vec}(\mathbf X_i)]$ is an $(rc) \times (rc)$ positive-definite covariance matrix,
\item $\mathbf Z_1,\ldots,\mathbf Z_N$ are i.i.d. $r \times c$ random matrices and $Z_{iab}$ is the $(a,b)$-th element of $\mathbf Z_i$,
\item $\mathrm{E}[Z_{iab}]=0$, $\mathrm{E}[Z^{2}_{iab}]=1$, $\mathrm{E}[Z^4_{iab}]=3+B$ for a finite constant $B>-2$, $\mathrm{E}[Z^8_{iab}]<\infty$ and for any positive integers $l_1,\ldots,l_q$ with $\sum_{\nu=1}^q l_{\nu} \leq 8$ 
\begin{equation*}
\mathrm{E}[Z^{l_1}_{ia_1b_1} Z^{l_2}_{ia_2b_2} \ldots Z^{l_q}_{ia_qb_q}]=\mathrm{E}[Z^{l_1}_{ia_1b_1}]\mathrm{E}[Z^{l_2}_{ia_2b_2}] \ldots \mathrm{E}[Z^{l_q}_{ia_qb_q}]
\end{equation*}
for $(a_1,b_1)\neq (a_2,b_2) \neq \cdots \neq (a_q,b_q)$. 
\end{enumerate}
The matrix-variate normal distribution \citep{Dawid1981,Gupta2000}, a common and sensible choice for modeling transposable data, is a special case of model~(\ref{Nonparametricmodel}). To see this, let $Z_{iab}$ be i.i.d. random variables from a standard normal distribution $\mathrm{N}(0,1)$ and let $\boldsymbol \Sigma=\boldsymbol \Sigma_2 \otimes \boldsymbol \Sigma_1$, where $\boldsymbol \Sigma_1$ is the covariance matrix of the row variables, $\boldsymbol \Sigma_2$ is the covariance matrix of the column variables and $\otimes$ denotes the Kronecker product operator applied to matrices. However, we underline that model~(\ref{Nonparametricmodel}) is more general. It can handle departures from the matrix-variate normal model by relaxing the normality and/or the covariance structure assumption. The distribution of the ``white-noise'' random variables in $\mathbf Z_i$ remains unspecified. In fact, the white noise random variables do not need to be independent or identically distributed. Also the dependence structure between and among the row and column variables is not limited to a Kronecker product form. 

\indent To construct the test statistic for testing hypothesis~(\ref{arbitrarygroups}), we need additional notation. Let $\mathbf P_{\{c_1,c_2,\ldots,c_g\}}= \mathrm{diag}(\mathbf P_{c_1},\mathbf P_{c_2},\ldots,\mathbf P_{c_g})$ be the $c \times c$ block diagonal matrix where the positive integers $\{c_1,c_2,\ldots,c_g\}$ are defined by $\mathrm{H}_0$ in~(\ref{arbitrarygroups}). For notational ease, suppress the index set in $\mathbf P_{\{c_1,c_2,\ldots,c_g\}}$ and write instead $\mathbf P$. Further, note that $\mathbf P$ is a projection matrix as it is both idempotent and symmetric. The key to our proposal is to observe that $\mathrm{tr}(\mathbf M^T \mathbf M \mathbf P)= 0$ if and only if $\mathrm{H}_0$ in~(\ref{arbitrarygroups}) holds. To see this, note that $\mathrm{tr}(\mathbf M^T \mathbf M \mathbf P)=\mathrm{tr}(\mathbf P \mathbf M^T \mathbf M \mathbf P)$ is the sum of squares of the elements of $\mathbf M \mathbf P$, whose $(a,b)$-th element equals the difference between $\mu_{ab}$, the $(a,b)$-th element of $\mathbf M$, and $\bar{\mu}^{(k)}_{a}$, the average of the $a$-th row in the mean matrix when this is restricted to the column group, say $k$, to which column $b$ belongs under $\mathrm{H}_0$ in~(\ref{arbitrarygroups}). Therefore, it is sensible to consider the unbiased estimator of $\mathrm{tr}(\mathbf M^T \mathbf M \mathbf P)$ 
$$G_N=\frac{1}{N(N-1)} \sum_{i \neq j} \mathrm{tr}(\mathbf X_i^{T} \mathbf X_j \mathbf P),$$
whose variance is
\begin{equation*}
\mathrm{Var}[G_N]=\frac{2}{N(N-1)}\mathrm{tr}\left([\boldsymbol \Sigma (\mathbf P \otimes \mathbf I_{r})]^2\right)+\frac{4}{N}\mathrm{vec}(\mathbf M \mathbf P)^{T} \boldsymbol \Sigma \mathrm{vec}(\mathbf M \mathbf P).
\end{equation*}

\indent Next, we define the asymptotic framework needed to derive the limiting distribution of $G_N$. We handle the high-dimensional setting without specifying the limiting rate of the pairwise ratios of the triplet $(N,r,c)$ because in many applications, including our motivating examples, the number of row (genes) and/or column (multiple samples) variables are not expected to increase proportionally to the sample size. Instead, we assume that as $N \rightarrow \infty$ and $rc=r(N)c(N) \rightarrow \infty$, the following conditions hold:
\begin{equation}
\mathrm{tr}\left([\boldsymbol \Sigma (\mathbf P \otimes \mathbf I_{r})]^4\right) =o\left\{\mathrm{tr}^2\left([\boldsymbol \Sigma (\mathbf P \otimes \mathbf I_{r})]^2\right) \right\}
\label{CovarianceAssumption}
\end{equation}
and 
\begin{equation}
\mathrm{vec}(\mathbf M \mathbf P)^{T} \boldsymbol \Sigma \mathrm{vec}(\mathbf M \mathbf P)=o\left\{\frac{1}{N}\mathrm{tr}\left([\boldsymbol \Sigma (\mathbf P \otimes \mathbf I_{r})]^2\right)\right\}
\label{MeanAssumption}
\end{equation}
or
\begin{equation}
\frac{1}{N}\mathrm{tr}\left([\boldsymbol \Sigma (\mathbf P \otimes \mathbf I_{r})]^2\right)=o\left\{\mathrm{vec}(\mathbf M \mathbf P)^{T} \boldsymbol \Sigma \mathrm{vec}(\mathbf M \mathbf P)\right\}.
\label{PowerAssumption}
\end{equation}
The assumption $rc \rightarrow \infty$ does not require $r \rightarrow \infty$ and $c \rightarrow \infty$ simultaneously and it allows the number of row or column variables to be fixed provided that the other dimension of the transposable data tends to $\infty$. Condition~(\ref{CovarianceAssumption}) specifies the class of covariance matrices for $\boldsymbol \Sigma$ under consideration. In Section~\ref{CovarianceClass}, we argue that this class is quite large and thus, the proposed testing procedure is not seriously restricted. At least one of the conditions~(\ref{MeanAssumption}) and~(\ref{PowerAssumption}) is needed to control the asymptotic variance of $G_N$ and to derive the asymptotic distribution of $G_N$, given in Theorem~\ref{MainTheorem} and proven in the Web Appendix A.
\begin{theorem}
Under the nonparametric model~(\ref{Nonparametricmodel}), condition~(\ref{CovarianceAssumption}) and either condition~(\ref{MeanAssumption}) or condition~(\ref{PowerAssumption})
$$\frac{G_N-\mathrm{tr}(\mathbf M^{T} \mathbf M \mathbf P)}{\sqrt{\mathrm{Var}[G_N]}} \leadsto \mathrm{N}(0,1)$$
where $\leadsto$ denotes convergence in distribution as $N \rightarrow \infty$ and $rc=r(N)c(N) \rightarrow \infty$. Consequently, under $\mathrm{H}_0$ in~(\ref{arbitrarygroups}), 
$$\frac{G_N}{\sqrt{2\mathrm{tr}\left([\boldsymbol \Sigma (\mathbf P \otimes \mathbf I_{r})]^2\right)/[N(N-1)]}} \leadsto \mathrm{N}(0,1).$$
\label{MainTheorem}
\end{theorem}
To construct the test statistic, we avoid estimating the unknown and high-dimensional covariance matrix $\boldsymbol \Sigma$ upon observing that the $N$ i.i.d. $rc$-variate random vectors $\mathbf Y_i=\mathrm{vec}(\mathbf X_i \mathbf P)$ have covariance matrix $\boldsymbol \Omega=(\mathbf P \otimes \mathbf I_{r}) \boldsymbol \Sigma (\mathbf P \otimes \mathbf I_{r})$ and that $\mathrm{tr}(\boldsymbol \Omega^2)=\mathrm{tr}\left([\boldsymbol \Sigma (\mathbf P \otimes \mathbf I_{r})]^2\right)$. Therefore, it follows from the work of \cite*{Chen2010a} that 
$$T_{N}=\frac{1}{D^N_2}\sum\nolimits_{i \neq j} (\mathbf Y^{T}_{i}\mathbf Y_{j})^2-2\frac{1}{D^N_3}\sum\nolimits_{i \neq j \neq k}^{\ast} \mathbf Y^{T}_{i}\mathbf Y_{j}\mathbf Y^{T}_{i}\mathbf Y_{k}+\frac{1}{D^N_4} \sum\nolimits_{i\neq j \neq k \neq l}^{\ast} \mathbf Y^{T}_{i}\mathbf Y_{j}\mathbf Y^{T}_{k}\mathbf Y_{l}$$ 
where $D^s_t=(s-t)!/s!$ and $\sum^{\ast}$ denotes summation over mutually exclusive indices, is a ratio-consistent estimator of $\mathrm{tr}(\boldsymbol \Omega^2)$. Therefore, the proposed test rejects $\mathrm{H}_0$ in (\ref{arbitrarygroups}) with an $\alpha$ significance level if and only if 
$$G^{\ast}_N=\frac{G_N}{\sqrt{2T_{N}/[N(N-1)]}} \geq z_a,$$ 
where $z_a$ is the upper $\alpha$-quantile of $\mathrm {N}(0,1)$.

\subsection{Remarks}
Consider the transformation $\mathbf X_i \longmapsto a \mathbf A \mathbf X_i + \mathbf C$ where $a \neq 0 \in \Re$, $\mathbf A$ is an $r \times r$ orthogonal matrix and $\mathbf C$ is an $r \times c$ matrix of constants such that $\mathbf C \mathbf P= \mathbf 0_{r \times c}$, and where $\mathbf 0_{s \times t}$ denotes the zero matrix of size $s \times t$. As desired, the test statistic $G^{\ast}_N$ is invariant to orthogonal rotations of the row variables, to scalar multiplication, and to location shifts of the mean matrix under $\mathrm{H}_0$ in (\ref{arbitrarygroups}). The last property implies that the nominal size of the test statistic is not affected by the magnitude of the true mean matrix $\mathbf M$ given that this satisfies $\mathrm{H}_0$ in (\ref{arbitrarygroups}). To this end, note that column groups of size one do not contribute to the test statistic, meaning that the value of $G^{\ast}_N$ does not change if column groups of size one ($c_k=1$) are ignored. This is not surprising since no mean comparisons are performed therein. Hence, these column variables should be removed prior to calculating the test statistic.  

\indent Although the testing methodology is presented for testing the mean structure of row variables across groups of column variables, we emphasize that the same testing procedure can be used to test the mean structure of column variables across groups of row variables. To do this, apply the transformation $\mathbf X_i \longmapsto \mathbf X^{T}_i$ prior to calculating $G^{\ast}_N$. 

\indent A critical point in our proposal is the choice of the projection matrix $\mathbf P$. Although Theorem~\ref{MainTheorem} holds for any projection matrix that satisfies the required assumptions, say $\mathbf P^{\ast}$, to avoid trivial power under certain alternatives it is essential to require that $\mathbf M \mathbf P^{\ast}=\mathbf 0_{r \times c}$ if and only if the corresponding null hypothesis is true. For example, an alternative way to test hypothesis~(\ref{onegroup}) is to consider the projection matrix $\mathbf P^{\ast}=\mathbf J_{c}/c$ (instead of $\mathbf P_c=\mathbf I_c - \mathbf J_{c}/c$). The asymptotic power of the resulting test statistic is trivial if, for example, $c$ is even and the mean vector is $\boldsymbol \mu$ for the odd columns of $\mathbf M$ and $-\boldsymbol \mu$ for the even columns. Thus attention is required when projection matrices other than the suggested ones are used.

\indent It is important to note that the testing procedure can be modified and applied to test hypotheses other than hypothesis~(\ref{arbitrarygroups}). For example, consider testing the hypothesis of a known $r \times c$ matrix of constants $\mathbf M_0$ ($\mathrm{H_0}: \mathbf M= \mathbf M_0$). To do this, we can center the data by subtracting $\mathbf M_0$ and then employ the test statistic $G^{\ast}_N$ calculated using $\mathbf P=\mathbf I_{c}$. Another example is testing the hypothesis $\mathrm{H_0}: \boldsymbol \mu_1-\boldsymbol \mu_2=\boldsymbol \mu_0$, where $\boldsymbol \mu_1$ and $\boldsymbol \mu_2$ are the unknown $r$-variate mean vectors of the first and second column variable respectively, and $\boldsymbol \mu_0$ is an $r$-variate vector of known constants. To accomplish this, one needs to subtract $\boldsymbol \mu_0$ from the first column of each data matrix and then test hypothesis~(\ref{arbitrarygroups}) with $g=2$, $c_1=2$ and $c_2=\ldots=c_g=1$ using the transformed data. In a similar way, the proposed method can be extended to test known differences in the mean vectors of two or more column groups. 

\indent To calculate $T_N$, it is more efficient to use the equivalent formula given in \cite{Himenoa2012} which reduces the computational cost from $O(N^4)$ to $O(N^2)$. Combining this result with simple algebraical properties for the trace operator, we can prove that the proposed testing methodology is computationally cheap regardless of the dimensionality, i.e., number of row variables, number of column variables or sample size.

\subsection{Asymptotic power}\label{PowerExamples}
Under condition~(\ref{MeanAssumption}), the leading order power for the proposed test is 
\begin{equation*}
\beta_{N}=\Phi\left(-z_a+N\frac{\mathrm{tr}(\mathbf M^{T} \mathbf M \mathbf P)}{\sqrt{2\mathrm{tr}(\boldsymbol \Omega^2)}}\right),
\end{equation*} 
where $\Phi$ is the cumulative distribution function of $\mathrm{N}(0,1)$. The power of the proposed test is bounded since
$$\Phi\left(-z_a+N\frac{\mathrm{tr}(\mathbf M^{T} \mathbf M \mathbf P)}{\sqrt{2\mathrm{tr}(\boldsymbol \Sigma^2)}}\right)\leq \beta_{N} \leq  \Phi\left(-z_a+N\frac{\mathrm{tr}(\mathbf M^{T} \mathbf M)}{\sqrt{2\mathrm{tr}(\boldsymbol \Omega^2)}}\right),$$ and thus a sufficient condition for the proposed test to have non-trivial power is 
\begin{equation*}
\lim_{N,(rc)\rightarrow \infty}N\frac{\mathrm{tr}(\mathbf M^{T} \mathbf M \mathbf P)}{\sqrt{2\mathrm{tr}(\boldsymbol \Sigma^2)}}>0.
\label{sufcon1}
\end{equation*}

\indent Under condition~(\ref{PowerAssumption}), the leading order power term becomes 
\begin{equation*}
\beta_{N}=\Phi\left(\frac{\sqrt{N}\mathrm{tr}{(\mathbf {M}^{T}}\mathbf {M}\mathbf P)}{2\sqrt{\mathrm{vec}(\mathbf {M}\mathbf P)^T\boldsymbol \Sigma \mathrm{vec}(\mathbf {M}\mathbf P)}}\right)=\Phi\left(\frac{\sqrt{N}\mathrm{tr}{(\mathbf {M}^{T}}\mathbf {M}\mathbf P)}{2\sqrt{\mathrm{vec}(\mathbf {M})^T\boldsymbol \Omega \mathrm{vec}(\mathbf {M})}}\right).
\end{equation*}
The power of the proposed test remains bounded since
$$\Phi\left(\frac{\sqrt{N}\mathrm{tr}(\mathbf M^{T} \mathbf M \mathbf P)}{2\sqrt{\mathrm{vec}(\mathbf {M})^T\boldsymbol \Sigma \mathrm{vec}(\mathbf {M})}}\right)\leq \beta_{N} \leq  \Phi\left(\frac{\sqrt{N}\mathrm{tr}(\mathbf M^{T} \mathbf M)}{2\sqrt{\mathrm{vec}(\mathbf {M})^T\boldsymbol \Omega \mathrm{vec}(\mathbf {M})}}\right),$$
which implies that 
\begin{equation*}
\lim_{N,(rc)\rightarrow \infty}\frac{\sqrt{N}\mathrm{tr}(\mathbf M^{T} \mathbf M \mathbf P)}{2\sqrt{\mathrm{vec}(\mathbf {M})^T\boldsymbol \Sigma \mathrm{vec}(\mathbf {M})}}>0
\label{sufcon2}
\end{equation*}
is a sufficient condition for the proposed test to have non-trivial power.
 
\indent Although the proposed testing procedure can handle dependence structures other than the independence, it can still be more powerful than typical univariate tests that require multiple testing corrections even for independent row and column variables ($\boldsymbol \Sigma=\mathbf I_{rc}$). To provide such an instance, assume a fixed number of column variables and no row-effect in the mean structure, that is $M_{ab}=M_{b}$ where $M_{ab}$ is the $(a,b)$-th element of $\mathbf M$. In this scenario, the asymptotic power of the proposed test under conditions~(\ref{MeanAssumption}) and~(\ref{PowerAssumption}) becomes
\begin{equation*}
\Phi\left(-z_a+\sqrt{\frac{N^{2}r}{2(c-g)}}\sum_{k=1}^{g}\sum_{b=c_{k-1}+1}^{c_{k}}(M_{b}-\bar{M}^{(k)})^2\right) \text{  and  } \Phi\left(\frac{r}{2}\sqrt{N\sum_{k=1}^{g}\sum_{b=c_{k-1}+1}^{c_{k}}(M_{b}-\bar{M}^{(k)})^2}\right)
\end{equation*} 
respectively, where $c_0=0$ and $\bar{M}^{(k)}$ is the average of the mean of the row variable $a$ in group $k$. As desired, under either~(\ref{MeanAssumption}) or~(\ref{PowerAssumption}), the power of the test is an increasing function of the number of row variables $r$. On the contrary, the power of some commonly used univariate tests applied sequentially to each row, such as ANOVA based tests, depends on the magnitude of the differences $\{M_{b}-\bar{M}^{(k)},b=1,\ldots,c\}$. Therefore, we expect ANOVA based tests to suffer from low power when these differences are small regardless of $r$. Note that we reach to the same conclusion even if we replace the no row-effect in the mean structure with an unstructured one such that all row-wise differences $\{M_{ab}-\bar{M}^{(ak)},b=1,\ldots,c\}$ are small, and where $\bar{M}^{(ak)}$ denotes the average of the mean of the row variable $a$ in group $k$. In these cases, the proposed test performs better because it extracts information from both the row and the column variables, which is ignored by univariate tests. We verified this speculation in simulations where we also investigated the situation in which the null hypothesis under consideration was violated for varying proportions of the rows in the mean matrix.

\subsection{Class of covariance matrices under consideration}\label{CovarianceClass}
We provide examples of covariance matrices that satisfy condition~(\ref{CovarianceAssumption}) and technical details can be found in Web Appendix C. Because of the popularity of the matrix-variate normal distribution in modelling transposable data, we first study the implications of condition~(\ref{CovarianceAssumption}) when $\boldsymbol \Sigma= \boldsymbol \Sigma_2 \otimes \boldsymbol \Sigma_1$. In this case, condition~(\ref{CovarianceAssumption}) becomes 
$$\mathrm{tr}\left[(\mathbf P \boldsymbol \Sigma_2)^4\right]\mathrm{tr}( \boldsymbol \Sigma_1^4)=o\left\{\mathrm{tr}^2\left[(\mathbf P \boldsymbol \Sigma_2)^2\right]\mathrm{tr}^2( \boldsymbol \Sigma_1^2)\right\}.$$ 
For example, this condition is met if $\mathrm{tr}\left[(\mathbf P \boldsymbol \Sigma_2)^4\right]=o\left\{\mathrm{tr}^2\left[(\mathbf P \boldsymbol \Sigma_2)^2\right]\right\}$ and/or if $\mathrm{tr}( \boldsymbol \Sigma_1^4)=o\left\{\mathrm{tr}^2( \boldsymbol \Sigma_1^2)\right\}$. This means that $\boldsymbol \Sigma_1$ and/or $\boldsymbol \Sigma_2$ can have bounded eigenvalues or a few eigenvalues that diverge slowly to infinity \citep{Chen2010}, or satisfy a (banded) first order autoregressive correlation pattern such that the corresponding variances are bounded away from $0$ or $\infty$ \citep{Chen2010a}. When $c$ is fixed, then condition~(\ref{CovarianceAssumption}) becomes $\mathrm{tr}(\boldsymbol \Sigma_1^4) =o\left\{\mathrm{tr}^2(\boldsymbol \Sigma_1^2)\right\}$, and it follows that $\boldsymbol \Sigma_1$ cannot satisfy a compound symmetry correlation structure. However, if $r$ is fixed, then condition~(\ref{CovarianceAssumption}) becomes $\mathrm{tr}\left[(\mathbf P \boldsymbol \Sigma_2)^4\right] =o\left\{\mathrm{tr}^2\left[(\mathbf P \boldsymbol \Sigma_2)^2\right]\right\}$, and therefore the compound symmetry correlation structure is an acceptable dependence structure for $\boldsymbol \Sigma_2$.  

\indent A sufficient assumption for condition~(\ref{CovarianceAssumption}) in the presence of uncorrelated (not necessarily independent) column variables is that $\mathrm{tr}(\boldsymbol \Sigma^4)=o\left\{\mathrm{tr}^2(\boldsymbol \Sigma^2) \right\}$. This assumption covers the case of independent row and column variables with bounded variances or a few divergent variances among others. When the row and column variables are correlated, then condition~(\ref{CovarianceAssumption}) is met for a covariance matrix $\boldsymbol \Sigma$ with bounded eigenvalues or a few divergent values that diverge slowly, for $\boldsymbol \Sigma$ that implies a (banded) first order autoregressive correlation pattern or a (banded) compound symmetry correlation matrix. 

\section{Simulation Studies}\label{Simulation}
We investigated the nominal size and the power of the proposed testing procedure using simulations. The simulated random matrices $\mathbf X_1, \ldots,\mathbf X_N$ satisfied model~(\ref{Nonparametricmodel}). To study the nonparametric nature of the proposed methodology, three distributional scenarios were considered for the elements of $\mathbf Z_i$: 
\begin{enumerate}
\item A normality scenario, in which $Z_{iab} \stackrel{i.i.d}{\sim} \mathrm{N}(0,1)$.
\item A centralized gamma distributional scenario, in which $Z_{iab}=(Z^{\ast}_{iab}-8)/4$ and $Z^{\ast}_{iab} \stackrel{i.i.d}{\sim} \mathrm{Gamma}(4,0.5)$.
\item A mixture of Scenarios 1 and 2, in which the random variables in the upper half of $\mathbf Z_i$ are distributed as in Scenario 1, while the remaining random variables are distributed as in Scenario 2.
\end{enumerate}
Conditional on $N$, $\mathbf M$, $\boldsymbol \Sigma$ and the distributional scenario, we draw 1000 replicates while keeping the significance level fixed at $5\%$. For each competing testing procedure, we calculated the empirical size as the proportion of rejections when $\mathbf M = \mathbf 0_{r \times c}$ and the empirical power as the proportion of rejections when $\mathbf M \neq \mathbf 0_{r \times c}$ as defined in Sections \ref{sectest1} and \ref{sectest3}. To distinguish the test statistics of the proposed methodology used in the simulations, we denoted by $H_{\{c_1,c_2,\ldots,c_g\}}$ the test statistic $G^{\ast}_N$ of the proposed methodology based on $\mathbf P_{\{c_1,c_2,\ldots,c_g\}}$. Further, we let $[k]$ denote the integer part of $k \in \Re$.  Additional simulation studies for the proposed testing methodology can be found on the Web Appendix B.

\subsection{Comparison with ANOVA and Kruskal-Wallis} \label{sectest1}
We first compared the proposed testing methodology, evaluated using $H_{\{c\}}$, to the ANOVA test of no group effect and the Kruskal-Wallis test for testing the hypothesis of no column effect in the mean matrix, i.e., testing hypothesis~(\ref{onegroup}). The ANOVA and Kruskal-Wallis tests were applied sequentially to each of the $r$ row variables and the resulting $p$-values were adjusted using the false discovery rate (FDR) correction and the Bonferroni (BON) correction. Web Table 2 suggests that the ANOVA and Kruskal-Wallis tests are extremely conservative in the presence of row-wise and column-wise dependencies and therefore, a fair and meaningful comparison is ensured by restricting the dependence structure to independent row and column variables ($\boldsymbol \Sigma=\mathbf I_{r c}$). In addition to calculating the empirical size, we measured the empirical power of the competing tests assuming that $\mathbf M= [\mathbf 0_{r \times 7} , t \mathbf J_{r \times 3}]$ where $\mathbf J_{k \times l}$ denotes the $k \times l$ matrix of ones. This configuration is motivated by the power analysis in Section~\ref{PowerExamples}. The constant $t$ was selected such that $\mathrm{tr}(\mathbf {M}^T \mathbf {M})/\sqrt{r(c-1)}=0.1$, i.e., by fixing the quantity that determines the upper bound of the asymptotic power of the proposed tests under condition~(\ref{MeanAssumption}) equal to $0.1$. In this way, the asymptotic power of $H_{\{c\}}$ is not trivial and the simulation results are comparable across varying values of $r$ and $c$. Table~\ref{tab1} displays the results under Scenario 3 - similar patterns were observed under the other two scenarios. Unlike the Kruskal-Wallis test which seemed to be conservative unless $N=100$, the empirical sizes for $H_{\{c\}}$ and for the ANOVA test appeared to be a good approximation of the nominal size even for $N=10$. Despite the conservativeness of the proposed test for $N=10$, it was always more powerful than the ANOVA and the Kruskal-Wallis test. Conditional on $N$ and the distributional scenario, the empirical power of the proposed test increased as $r$ increased while that of the competing testing procedures did not change much even when $N=100$. This is due to the effectiveness of the proposed test in high-dimensional settings when the magnitude of the row-wise (column-wise) difference in the mean matrix is small but constant for every row (column) of the mean structure.

\begin{table}
\centering
\caption{Empirical size and power of $H_{\{10\}}$, ANOVA and Kruskal-Wallis test at $5\%$ significance.}
\begin{tabular*}{\textwidth}{c @{\extracolsep{\fill}}cccccccccccc}
  \toprule
    & & & & \multicolumn{4}{c}{ANOVA}   & \multicolumn{4}{c}{Kruskal-Wallis} \\ 
    & & \multicolumn{2}{c}{$H_{\{10\}}$} & \multicolumn{2}{c}{FDR}    & \multicolumn{2}{c}{BON} & \multicolumn{2}{c}{FDR} & \multicolumn{2}{c}{BON}\\
 \toprule
 $r$ &$N$ & Power & Size  & Power & Size  & Power & Size  & Power & Size  & Power & Size \\ 
 \toprule
100 &10  & 0.138 & 0.063 & 0.051 & 0.047 & 0.051 & 0.046 & 0.013 & 0.014 & 0.013 & 0.014 \\ 
    &30  & 0.412 & 0.057 & 0.091 & 0.045 & 0.088 & 0.045 & 0.062 & 0.040 & 0.060 & 0.039 \\ 
    &50  & 0.756 & 0.053 & 0.136 & 0.045 & 0.125 & 0.044 & 0.115 & 0.043 & 0.112 & 0.043 \\ 
    &100 & 0.997 & 0.044 & 0.319 & 0.047 & 0.294 & 0.045 & 0.317 & 0.048 & 0.285 & 0.047 \\ 
500 &10  & 0.186 & 0.063 & 0.075 & 0.066 & 0.075 & 0.066 & 0.011 & 0.008 & 0.011 & 0.008 \\ 
    &30  & 0.703 & 0.039 & 0.096 & 0.060 & 0.094 & 0.059 & 0.051 & 0.033 & 0.047 & 0.033 \\ 
    &50  & 0.974 & 0.040 & 0.102 & 0.042 & 0.093 & 0.040 & 0.082 & 0.026 & 0.077 & 0.026 \\ 
    &100 & 1.000 & 0.051 & 0.261 & 0.054 & 0.244 & 0.053 & 0.253 & 0.048 & 0.233 & 0.047 \\ 
   \bottomrule
\end{tabular*}
\label{tab1}
\end{table}

\indent Next, we compared the empirical power of the competing testing procedures under a sparsity scenario for the mean structure. In particular, we defined $\mathbf M = [\mathbf 0_{r \times 9}, \boldsymbol \mu]$ and similarly to \cite{Chen2010}, we let the $r$-variate vector $\boldsymbol \mu$ contain a varying proportion ($0\%$, $25\%$, $50\%$, $75\%$, $95\%$ and $99\%$) of zero elements. At each proportion level, we employed a linearly increasing allocation where two nonzero-elements of $\boldsymbol \mu$ satisfy $\mu_{l_1} < \mu_{l_2}$ if and only if $l_1<l_2$. We set $r=100,500,1000$ and we let $\boldsymbol \Sigma=\mathbf I_{10r}$. To make the results comparable across the sampling schemes, the non-zero elements of $\boldsymbol \mu$ were defined in such a way that
$$\frac{\mathrm{tr}(\mathbf M^{T}\mathbf M)}{\sqrt{r (c-1)}}= 0.15.$$
Table~\ref{tab3} displays the simulation results only for $r=1000$ under Scenario 3 since similar trends were noted for the remaining sampling schemes. As desired, the empirical power of the proposed methodology appeared to be unaffected by the proportion of zero elements in $\boldsymbol \mu$ for fixed $N$ and the empirical power approached $1.00$ as soon as $N=50$. However, the empirical power of the ANOVA and Kruskal-Wallis tests seemed to decrease as the proportion of zero elements decreased. In fact, the largest differences between the empirical power of the proposed test and of the univariate tests were observed when there were no zeros in $\boldsymbol \mu$. This agrees with our claims in Section~\ref{PowerExamples} regarding the power of the competing procedures. For $1\%$ of non-zero elements in $\boldsymbol \mu$, the empirical powers of the three testing procedures were comparable unless $N=30$ in which case the ANOVA and Kruskal-Wallis tests were substantially more powerful than the proposed test. For all other proportions of zero elements in $\boldsymbol \mu$, the proposed test was extremely more powerful than the univariate tests with the sole exception of the sampling scheme with $N=100$ and $75\%$ of zero elements in $\boldsymbol \mu$. Overall, the proposed test appeared to be more powerful than univariate tests under the sparsity scenario for the mean matrix and under the rather unrealistic assumption of independent row and column variables for the dependence structure. Similar trends were observed for an equal allocation scenario in $\boldsymbol \mu$ (see Web Table 3).

\begin{table}
\centering
\caption{Empirical power of $H_{\{10\}}$, ANOVA and Kruskal-Wallis for the sparsity scenario with $r=1000$ under Scenario 3 at $5\%$ significance}
\begin{tabular*}{\textwidth}{c @{\extracolsep{\fill}}lcccccc}
  \toprule
   &   & $H_{\{10\}}$ & \multicolumn{2}{c}{ANOVA}  &  \multicolumn{2}{c}{Kruskal-Wallis} \\  
$N$& $\#\{\mu_{l}=0\}$ &   & FDR & BON & FDR & BON \\ 
\hline
10 & 99\% & 0.164 & 0.189 & 0.184 & 0.014 & 0.014 \\  
   & 95\% & 0.170 & 0.068 & 0.067 & 0.003 & 0.003 \\
   & 75\% & 0.162 & 0.062 & 0.061 & 0.003 & 0.003 \\ 
   & 50\% & 0.164 & 0.061 & 0.060 & 0.003 & 0.003 \\ 
   & 25\% & 0.161 & 0.061 & 0.060 & 0.004 & 0.004 \\
   & 0 \% & 0.168 & 0.058 & 0.057 & 0.003 & 0.003 \\ 
30 & 99\% & 0.618 & 0.997 & 0.997 & 0.976 & 0.971 \\ 
   & 95\% & 0.624 & 0.254 & 0.242 & 0.132 & 0.125 \\ 
   & 75\% & 0.618 & 0.096 & 0.091 & 0.052 & 0.050 \\ 
   & 50\% & 0.626 & 0.082 & 0.080 & 0.044 & 0.043 \\  
   & 25\% & 0.628 & 0.081 & 0.078 & 0.047 & 0.045 \\ 
   & 0 \% & 0.625 & 0.084 & 0.081 & 0.051 & 0.049 \\  
50 & 99\% & 0.949 & 1.000 & 1.000 & 1.000 & 1.000 \\  
   & 95\% & 0.948 & 0.721 & 0.678 & 0.566 & 0.538 \\ 
   & 75\% & 0.949 & 0.144 & 0.135 & 0.103 & 0.100 \\ 
   & 50\% & 0.943 & 0.117 & 0.108 & 0.080 & 0.078 \\ 
   & 25\% & 0.944 & 0.105 & 0.102 & 0.078 & 0.077 \\ 
   & 0 \% & 0.944 & 0.094 & 0.092 & 0.076 & 0.073 \\ 
100& 99\% & 1.000 & 1.000 & 1.000 & 1.000 & 1.000 \\ 
   & 95\% & 1.000 & 1.000 & 1.000 & 1.000 & 0.999 \\ 
   & 75\% & 1.000 & 0.398 & 0.356 & 0.314 & 0.290 \\
   & 50\% & 1.000 & 0.245 & 0.229 & 0.192 & 0.176 \\ 
   & 25\% & 1.000 & 0.197 & 0.180 & 0.157 & 0.148 \\ 
   & 0 \% & 1.000 & 0.163 & 0.152 & 0.155 & 0.142 \\ 
   \bottomrule
\end{tabular*}
\label{tab3}
\end{table}

\subsection{Comparison with the Chen-Qin test}
Suppose we want to test hypothesis~(\ref{onegroup}) when the column variables are independent. In this case, we can create $c$ groups, one group for each column variable that contains $N$ independent $r$-variate random vectors. An alternative practical approach to test hypothesis~(\ref{onegroup}) is to apply the two-sample test for high-dimensional mean vectors proposed by \cite{Chen2010} to all possible pairs of groups, and then adjust the resulting $p$-values for multiple testing. To satisfy the required assumptions of the Chen-Qin test, $\boldsymbol \Sigma$ was set equal to a block diagonal matrix with $c$ blocks. Each block of $\boldsymbol \Sigma$ satisfied a first-order autoregressive form ($\{\rho^{|a-b|}\}_{1\leq a,b \leq r}$) where $\rho=0.5$ in the first $c/2$ blocks and $\rho=0.4$ elsewhere. Table~\ref{tab5} shows the empirical sizes of the two competing testing procedures across the three distributional scenarios with $c=10$. The proposed test seemed to preserve the nominal size but the Chen-Qin test appeared to have a highly inflated empirical size even when $r=1000$, which prohibited us from conducting power comparisons. 

\begin{table}
\centering
\caption{Empirical size of $H_{\{10\}}$ and the Chen-Qin test (with a Bonferroni correction) at $5\%$ significance.}
\begin{tabular*}{\textwidth}{c @{\extracolsep{\fill}}cccccccc}
  \toprule
& & \multicolumn{2}{c}{Scenario 1} & \multicolumn{2}{c}{Scenario 2}  & \multicolumn{2}{c}{Scenario 3}\\ 
$r$ &$N$ & $H_{\{10\}}$ & Chen-Qin & $H_{\{10\}}$ & Chen-Qin & $H_{\{10\}}$ & Chen-Qin \\ 
  \toprule
 100 &10 & 0.048 & 0.179 & 0.066 & 0.179 & 0.065 & 0.173 \\ 
     &20 & 0.050 & 0.144 & 0.058 & 0.150 & 0.059 & 0.144 \\ 
     &30 & 0.059 & 0.147 & 0.069 & 0.157 & 0.056 & 0.158 \\ 
     &50 & 0.057 & 0.142 & 0.046 & 0.126 & 0.063 & 0.169 \\ 
 500 &10 & 0.045 & 0.114 & 0.059 & 0.104 & 0.057 & 0.097 \\ 
     &20 & 0.051 & 0.115 & 0.046 & 0.090 & 0.054 & 0.091 \\ 
     &30 & 0.054 & 0.084 & 0.046 & 0.081 & 0.040 & 0.078 \\ 
     &50 & 0.054 & 0.091 & 0.050 & 0.090 & 0.050 & 0.077 \\ 
1000 &10 & 0.060 & 0.093 & 0.051 & 0.081 & 0.057 & 0.087 \\ 
     &20 & 0.053 & 0.080 & 0.059 & 0.068 & 0.046 & 0.090 \\  
     &30 & 0.046 & 0.068 & 0.059 & 0.089 & 0.061 & 0.073 \\  
     &50 & 0.042 & 0.067 & 0.051 & 0.075 & 0.052 & 0.067 \\   
  \bottomrule
\end{tabular*}
\label{tab5}
\end{table}

\subsection{Nominal size}\label{sectest2}
Using $H_{\{c\}}$, $H_{\{[0.7 c],[0.3 c]\}}$ and $H_{\{[0.5 c],[0.2 c],[0.3 c]\}}$, we examined in greater detail the size of the proposed methodology with non-independence dependence patterns. In particular, we assumed that $\boldsymbol \Sigma=\boldsymbol \Sigma_2 \otimes \boldsymbol \Sigma_1$ where $\boldsymbol \Sigma_1=\{0.85^{|a-b|}\}_{1\leq a,b \leq r}$ and $\boldsymbol \Sigma_2=0.5(\mathbf I_{c}+\mathbf J_{c})$ and we employed an exchangeable form for $\boldsymbol \Sigma$ but since the results were similar, we present only the simulations with the Kronecker product dependence structure. To reflect practical situations where the dimension of the mean vector is at least equal to the sample size ($N$) and the number of row variables ($r$) is greater or equal to the number of column variables ($c$), we set $N=10,30,50,100$, $r=100,500,1000$ and $c=10,100$. Also, we covered the case where the number of row variables is much smaller than the number of column variables by using $r=10$ and $c=100,500$. Table~\ref{tab2} contains the empirical sizes under Scenario 3. Again, similar results were observed for the other two distributional scenarios, a fact that validates empirically the non-parametric nature of the methodology. The discrepancy between the empirical and nominal size was small for all three test statistics which confirms the robustness of the proposed testing procedure to the number of groups and to the group sizes.

\begin{table}
\centering
\caption{Empirical size of the proposed methodology under Scenario 3 and a Kronecker product dependence structure at $5\%$ significance.}
\begin{tabular*}{\textwidth}{c @{\extracolsep{\fill}}cccccccc}
   \toprule
$N$   & $r$   &  \multicolumn{2}{c}{$H_{\{c\}}$} & \multicolumn{2}{c}{$H_{\{[0.7 c],[0.3 c]\}}$}  & \multicolumn{2}{c}{$H_{\{[0.5 c],[0.2 c],[0.3 c]\}}$}\\ 
 \toprule
 &$c$ & 10    & 100   & 10    & 100   & 10    & 100 \\   
\cline{2-8}
10  & 100 & 0.064 & 0.056 & 0.059 & 0.056 & 0.057 & 0.058 \\ 
    & 500 & 0.068 & 0.067 & 0.068 & 0.067 & 0.060 & 0.067 \\ 
	  & 1000& 0.058 & 0.065 & 0.060 & 0.057 & 0.060 & 0.060 \\
30  & 100 & 0.063 & 0.053 & 0.061 & 0.050 & 0.060 & 0.049 \\ 
    & 500 & 0.049 & 0.054 & 0.053 & 0.048 & 0.049 & 0.049 \\
		& 1000& 0.049 & 0.057 & 0.048 & 0.063 & 0.056 & 0.056 \\
50  & 100 & 0.058 & 0.046 & 0.059 & 0.048 & 0.064 & 0.048 \\ 
    & 500 & 0.060 & 0.058 & 0.066 & 0.062 & 0.054 & 0.059 \\
		& 1000& 0.047 & 0.044 & 0.047 & 0.042 & 0.039 & 0.045 \\
100 & 100 & 0.047 & 0.055 & 0.050 & 0.053 & 0.057 & 0.058 \\ 
    & 500 & 0.047 & 0.048 & 0.049 & 0.040 & 0.048 & 0.044 \\
		& 1000& 0.051 & 0.068 & 0.055 & 0.068 & 0.051 & 0.067 \\
   \toprule
 &$c$ & 100    & 500   & 100    & 500   & 100  & 500 \\   
\cline{2-8}
10  &10& 0.055 & 0.065 & 0.052 & 0.067 & 0.052 & 0.068 \\ 
30  &10& 0.061 & 0.059 & 0.057 & 0.058 & 0.057 & 0.055 \\ 
50  &10& 0.054 & 0.053 & 0.057 & 0.053 & 0.056 & 0.054 \\ 
100 &10& 0.062 & 0.045 & 0.065 & 0.045 & 0.058 & 0.045 \\    
 \bottomrule   
\end{tabular*}
\label{tab2}
\end{table}

\subsection{Power considerations}\label{sectest3}
Using $H_{\{c\}}$, $H_{\{[0.6 c],[0.4 c]\}}$ and $H_{\{[0.4 c],[0.2 c],[0.4 c]\}}$, we also evaluated the empirical power of the proposed methodology under a multiplicative mean vectors scenario. In particular, we let $\mathbf M = [\mathbf J_{r \times [0.9c]}, t \mathbf J_{r \times [0.1c]}]$, where $t=1.15$, $\boldsymbol \Sigma_1=\{0.85^{|a-b|}\}_{1\leq a,b \leq r}$ and $\boldsymbol \Sigma_2=0.5(\mathbf I_{c}+\mathbf J_{c})$ for $r=100,500,1000$ and $c=10,100$. Table~\ref{tab4} displays the simulation results based on $H_{\{c\}}$ across the three distributional scenarios. The tests based on $H_{\{[0.6 c],[0.4 c]\}}$ and $H_{\{[0.4 c],[0.2 c],[0.4 c]\}}$ were more powerful and hence we do not show these results. Conditional on $N$, $r$ and $c$, the empirical power was similar across the three distributional scenario and, as desired, it approached $1.00$ as the sample size, the number of row or column variables increased.

\begin{table}
\centering
\caption{Empirical power of $H_{\{c\}}$ for the multiplicity scenario at $5\%$ significance.}
\begin{tabular*}{\textwidth}{c @{\extracolsep{\fill}} cccccccc}
  \toprule
 & $c$      & 10 & 100 & 10 & 100 & 10 & 100 \\ 
 \toprule
 $N$ & $r$            & \multicolumn{2}{c}{Scenario 1} & \multicolumn{2}{c}{Scenario 2}  & \multicolumn{2}{c}{Scenario 3}\\   
   \toprule
10 & 100 & 0.097 & 0.317 & 0.128 & 0.282 & 0.103 & 0.303 \\ 
   & 500 & 0.210 & 0.778 & 0.207 & 0.813 & 0.206 & 0.781 \\ 
	 &1000 & 0.331 & 0.967 & 0.305 & 0.971 & 0.315 & 0.965 \\ 
30 & 100 & 0.291 & 0.975 & 0.313 & 0.964 & 0.294 & 0.966 \\ 
   & 500 & 0.809 & 1.000 & 0.782 & 1.000 & 0.790 & 1.000 \\ 
	 &1000 & 0.979 & 1.000 & 0.965 & 1.000 & 0.971 & 1.000 \\ 
50 & 100 & 0.590 & 1.000 & 0.551 & 1.000 & 0.576 & 1.000 \\ 
   & 500 & 0.997 & 1.000 & 0.992 & 1.000 & 0.998 & 1.000 \\ 
	 &1000 & 1.000 & 1.000 & 1.000 & 1.000 & 1.000 & 1.000 \\ 
   \bottomrule
\end{tabular*}
\label{tab4}
\end{table}

\section{Two Examples}\label{Example}
We applied the proposed testing methodology to two datasets.
 
\subsection{The glioblastoma dataset}\label{GBexample}
The glioblastoma (GB) dataset describes an experimental study designed to explore the heterogeneity of GB \citep{Sottoriva} by comparing the gene expression patterns in $3$ different brain compartments; the tumor margin (MA), normal brain tissue that surrounds the tumor mass, the subventricular zone (SVZ), a targeted area located at the center of the brain, and the tumor mass. For each of the patients ($N=8$) included in the study, $7$ mRNA samples were extracted; $1$ from the MA, $1$ from the SVZ and $5$ from different fragments in the tumor mass such that earlier fragments were closer to MA and later fragments closer to SVZ. Gene expression levels were then measured from the $7 \times 8 = 56$ mRNA samples using microarrays. The data for each subject were organized in a matrix with row variables ($r=16810$) the genes and column variables ($c=7$) the MA, the SVZ and the $5$ tumor fragments ordered in the spatial order described above.

\indent An important biological hypothesis was the conservation of the mean vectors of gene expression levels across the tumor mass. Statistically speaking, this corresponds to testing the hypothesis 
\begin{equation}
\mathrm{H}_0: \mathbf M =[\boldsymbol \mu_1,\boldsymbol \mu_2, \boldsymbol \mu_3 \mathbf 1^{T}_{5}] \text{ vs. } \mathrm{H}_1: \text{ not } \mathrm{H}_0,
\label{GBhyp1}
\end{equation}
where $\boldsymbol \mu_1$ and $\boldsymbol \mu_2$ denote the mean vector of gene expression levels in the MA and the SVZ respectively, and $\boldsymbol \mu_3$ denotes the common mean vector of gene expression levels in each of the $5$ tumor fragments. The corresponding test statistic was equal to $-0.282$ ($p$-value$=0.611$) suggesting that we did not have enough evidence to reject $\mathrm{H}_0$ in~(\ref{GBhyp1}). This motivated us to assess the likelihood of a simpler mean structure than the one tested in~(\ref{GBhyp1}) (see Web Table 1). These results suggest that the overall gene expression patterns differed across the $3$ brain compartments under study and thus, $\mathbf M =[\boldsymbol \mu_1,\boldsymbol \mu_2, \boldsymbol \mu_3 \mathbf 1^{T}_{5}]$ described adequately the compartment-wise mean relationship in the GB dataset.

\indent We compared further the mean gene expression patterns in the MA and the tumor mass by utilizing Gene Ontology (GO) terms. The GO terms classify genes into groups such that the genes within a group are involved in the same biological process. From the $1316$ gene groups in the GB dataset, we selected $231$ groups that had more than $7$ genes in order to be closer to the high-dimensional assumptions. For the $k$-th group of genes $(k=1,\ldots,231)$ with mean matrix $\mathbf M_k$, we tested the hypothesis 
\begin{equation*}
\mathrm{H}_{0k}: \mathbf M_k =[\boldsymbol \mu_{1k},\boldsymbol \mu_{2k}, \boldsymbol \mu_{1k} \mathbf 1^{T}_{5}] \text{ vs. } \mathrm{H}_{1k}: \text{ not } \mathrm{H}_{0k},
\end{equation*}
where $\boldsymbol \mu_{1k}$ denotes the common mean gene expression levels vector in the MA and in the $5$ tumor fragments, and $\boldsymbol \mu_{2k}$ denotes the mean gene expression levels vector in the SVZ. After applying an FDR correction, we rejected the null hypothesis in $224$ groups. The high-proportion of rejections ($96.97\%$) supports the adopted form for the overall mean matrix $\mathbf M$. Many of these $224$ gene-groups correspond to biological processes that are known to be directly linked to cancer, including cellular response to hypoxia and the extracellular matrix organization \citep{Gilkes2014}, negative regulation of retinoic acid receptor signaling pathway \citep{Tang2011,Connolly2013} and positive regulation of ERK1 and ERK2 cascade \citep{Santamaria2010} among others. Thus, rejecting the corresponding $\mathrm{H}_{0k}$ can be biologically justified.

\subsection{The mouse aging dataset}
The atlas of gene expression in the mouse aging data \citep{Zahn2007} contains mouse mRNA gene expression levels measured in different tissues. For each mouse ($N=40$), mRNA expression levels were extracted for $r=8932$ genes from up to $16$ tissues. Here, we considered $c=9$ tissues (adrenal glands, cerebrum, hippocampus, kidney, lung, muscle, spinal cord, spleen and thymus) for which mRNA gene expression levels were available for all the mice. 

\indent Unsurprisingly, the hypothesis of no tissue effect upon the mean expression level was rejected since $G^{\ast}_N=481.28$ ($p$-value$<0.001$). A subset of genes called `housekeeping' genes are typically assumed to be expressed at a relatively constant level across many or all known experimental conditions. As a result, these genes are often used to calibrate gene expression levels across experiments. However, it has been suggested that commonly used housekeeping genes can show considerable variability in expression across tissues \citep{Jonge2007,Kouadjo2007}. To explore this, we created a list of $22$ housekeeping genes compromised of $8$ genes that are commonly classified as housekeeping genes \citep{Jonge2007} and $14$ genes that were classified as housekeeping genes by \cite{Jonge2007}. The hypothesis of conservation of the mean expression levels of this gene-set across the $9$ tissues was rejected ($G^{\ast}_N=382.93$ and $p$-value$<0.001$). We believe that further research is required in order to identify housekeeping genes for these $9$ tissues and the proposed testing methodology is a useful statistical tool to this direction.

\section{Discussion}\label{Discussion}
We proposed a novel non-parametric procedure to test the mean matrix in high-dimensional transposable data. In particular, our methodology can determine whether in each of the given groups of column variables the mean of every row variable remains constant. Of course, the role of the row and column variables is interchangeable in transposable data and hence the proposed tests can be applied to check the effect of the row variables upon the mean vector of the column variables. The simulation studies verified the robustness of the proposed testing procedure to the number of row or column groups, to the size of each group, to the number of column and row variables relative to the sample size, and to the underlying dependence structure between and among the row and column variables. In simulations, the proposed tests were more powerful than univariate testing procedures that require row-wise and/or column-wise independence in almost all settings. In a sense, we developed a theoretically sound non-parametric testing procedure that extends the application of univariate ANOVA flavored tests to high-dimensional transposable data while making mild dependence structure assumptions. The practical advantage of the proposed test is its computationally simplicity since the cumbersome task of estimating high-dimensional matrix parameters, such as the mean matrix and the covariance matrix, is avoided. The proposed testing methodology is implemented in the function \textit{meanmat.ts()} of the R package HDTD (aavailable at \href{{http://www.bioconductor.org/packages/3.0/bioc/html/HDTD.html}}{{http://www.bioconductor.org/packages/3.0/bioc/html/HDTD.html}}).  

\indent In practice, we expect that the experimental design will dictate the null hypothesis of interest about the mean-relationship between the row and column variables, as was the case with the glioblastoma dataset. In applications where it is not clear which column (or row) groups should be formed under the null hypothesis, the following strategy that can be helpful in determining the column-wise (row-wise) structure. First, test whether there is no column (row) effect upon the mean of the row (column) variables. If we fail to reject this hypothesis, assume that the mean of the row (column) variables is independent of the column (row) variables. Otherwise, perform the test that two column (row) variables have the same mean vector for all pairs of column (row) variables, and then adjust for multiple testing using an FDR or a Bonferroni correction. If all the adjusted $p$-values are very small, then assume an unstructured mean matrix $\mathbf M$ or transpose the data and repeat the above procedure for the column (row) variables. Otherwise, record the column (row) pairs for which the adjusted $p$-values$<0.05$, form $g$ column (row) groups and test hypothesis~(\ref{arbitrarygroups}) as this is determined by the $g$ groups.

\indent In future work, we aim to develop test statistics for hypotheses that cannot be directly handled by the proposed testing methodology, e.g. the hypothesis of a mean-restricted matrix \citep{Allen2010}, that is $\mathbf M= \boldsymbol \mu \mathbf 1^{T}_c +\mathbf 1_r \boldsymbol \nu^{T}$ where $\boldsymbol \mu$ is an $r$-variate vector of constants and $\boldsymbol \nu$ is a $c$-variate vector of constants, and hypotheses of testing simultaneously the presence of predefined row and column groups.

\section*{Acknowledgements}
We thank Inmaculada Spiteri for helpful comments on the example in Section~\ref{GBexample}.






\bibliographystyle{plainnat}
\bibliography{dissrefer.bib}




\label{lastpage}
\end{document}